\documentclass[nolineno]{aa}
\makeatletter
\@ifpackageloaded{linenoaa}{%
  \renewcommand{\linenumbers}{}
  \renewcommand{\nolinenumbers}{}
  \renewcommand{\modulolinenumbers}[1]{}
  
  \renewcommand{\runninglinenumbers}{}
  \renewcommand{\linenomath}{}
  \renewcommand{\endlinenomath}{}
  \def\@linenumber{}
  \def\thelinenumber{}
}{}
\makeatother
\AtBeginDocument{\let\linenumbers\relax}

\usepackage{hyperref}
\usepackage{graphicx}
\usepackage{txfonts}
\usepackage{amsmath}
\usepackage{bm}
\usepackage{wasysym}

\def \rsun {R\textsubscript{\astrosun}}
\def \kms {{\rm km\;s$^{-1}$}}

\def \mss {{\rm m\;s$^{-2}$}}
\def \deg {$^\circ$ }

\usepackage{color}
\begin{document}

\title{Statistics of blob properties in two types of coronal streamers}
\titlerunning{Blobs in two types of coronal streamers}
\authorrunning{Li H., Huang Z., et al.}
\author{Haiyi Li\inst{1} \and Zhenghua Huang\inst{2,3}\fnmsep\thanks{Corresponding author: huangzh@nju.edu.cn} \and Maria S. Madjarska\inst{4}\fnmsep\inst{5} \and Youqian Qi\inst{1} \and Hui Fu\inst{1} \and Ming Xiong\inst{6} \and Lidong Xia\inst{1}}
   \institute{Shandong Key Laboratory of Space Environment and Exploration Technology, Institute of Space Sciences, School of Space Science and Technology, Shandong University, Shandong, China
   \and Institute of Science and Technology for Deep Space Exploration, Suzhou Campus, Nanjing University, Suzhou 215163, China
   \and State Key Laboratory of Lunar and Planetary Sciences, Macau University of Science and Technology, Macau, China
   \and Max Planck Institute for Solar System Research, Justus-von-Liebig-Weg 3, 37077, G\"ottingen, Germany
   \and Space Research and Technology Institute, Bulgarian Academy of Sciences, Acad. G. Bonchev Str., Bl. 1, 1113, Sofia, Bulgaria
   \and State Key Laboratory of Solar Activity and Space Weather, National Space Science Center, Chinese Academy of Sciences, Beijing, 100049, China\\
}
\date{January 2026}

\abstract
{Previous studies have shown that a streamer blob might originate in the lower corona and thus be affected by activity in that region.
While the base of one streamer might differ from that of another, it can be cataloged into two distinct types: active region streamers (ARSs) that have active regions at their base, and quiet equatorial streamers (QESs) that do not have an active region underneath.
The difference between the blob properties in ARSs and those in QESs remains unknown.}
{We compare the properties of propagating blobs in ARSs and QESs.}
{By analyzing the whole-year observations from SOHO/LASCO/C2 in 2018, we carried out a statistical analysis of the properties of propagating blobs in ARSs and QESs.}
{We found that the properties of streamer blobs are very different from one blob to another.
The occurrence rate of blobs in ARSs is about twice as high as that in QESs.
On average, the ARS blobs have significantly higher initial velocities and slightly higher accelerations, 
but slightly lower heights of first appearance than the QES blobs.
There is a weak positive correlation between the initial velocities and heights of first appearance in the two groups of streamer blobs.
The correlation between the accelerations and heights of first appearance in ARS blobs is negative, while that in QES blobs is positive.}
{Our results provide statistical evidence that a higher degree of activity at the coronal base of a streamer 
can cause more dynamic blobs higher up, and that it affects the structures of the solar wind originating in the region.}

\maketitle
\thispagestyle{empty}

\section{Introduction} 
\label{sec_intro}
Coronal streamers are large-scale helmet structures with ray features extending into the interplanetary space for many solar radii\,\citep[e.g.][]{1973SoPh...31..105S,2009SoPh..258..243A,2011ApJ...734..114P,2015ApJ...800...90P,2020ApJ...895..123B,2023MNRAS.518.1776L}.
They are persistent in the solar corona at anytime during the solar minimum or maximum\,\citep{2007A&A...475..707S,2005A&A...430..701V,2025A&A...697A.155C}.
They are thought to be consequences of the complex interaction between the ambient solar wind 
and large-scale solar magnetic fields involving current sheets\,\citep{1992ApJ...392..310W,1992SSRv...61..393K}.

\par
Notably, the element abundance in streamer legs was found to be consistent with the abundances of the slow solar wind \citep{1995SSRv...72...49G}, 
further reinforcing the hypothesis of a close connection between streamers and slow solar wind \cite[e.g.,][]{1981JGR....86.5438G}.
Such a connection has recently been confirmed by observations of the Solar Orbiter/Metis coronagraph that were studied by \,\citet{2023PhPl...30b2905A}.
The authors found that the slow solar wind is distributed along the center of the coronal streamer current sheet, 
exhibiting the lowest velocity, but the highest acceleration rate. Quantitatively, 
the solar wind velocity along the center of the coronal streamer was approximately 115--117\,\kms at 4.0\,\rsun, and increased to 150--190\,\kms at 6.8\,\rsun.
Additionally, the variability in the slow solar wind abundances has been attributed to differences between individual streamers.

\par 
A supplementary contribution to the streamer-derived slow solar wind comes from blobs, which are transient structures that form at the cusps of helmet streamers\,\citep{1997ApJ...484..472S,1998ApJ...498L.165W,2009SoPh..258..129S,2009ApJ...694.1471S,2012SoPh..276..261S,2023A&A...672A.100L}.
Blobs are now understood to be flux ropes, forming when outward-expanding helmet streamer loops reconnect and/or pinch off 
\citep[e.g.][]{2000GeoRL..27..149W,2009ApJ...691.1936C,2009ApJ...694.1471S,2020ApJ...905..139L,2022A&A...659A.110R}.
Early works by \citet{1997ApJ...484..472S} characterized small-scale blob-like features, 
noting their relatively small initial size, low intensity, radial motion, slowly increasing speed, and location within the streamer belt. 
It was concluded that these features can passively trace the slow solar wind outflow. 
Later, \citet{2010ApJ...715..300S} established a clear correlation between 
coronal observations of small-scale solar wind transient structures and their in situ detections near 1~au, 
providing deeper insights into the evolution of such transient structures in the heliosphere. 
\citet{2025ApJ...988..152L} extended this work by tracing streamer blobs from several solar radii to the heliosphere using  the Solar Terrestrial Relations Observatory (STEREO) observations. 
They confirmed that the characteristics of the observed in situ properties (e.g., plasma density, temperature, and sheared magnetic polarity) 
and the surrounding plasma flow are fully consistent with the properties of the source regions of streamer blobs.

\par
Based on special coordinated observations from the C2 coronagraph of the Large Angle and Spectrometric Coronagraph\,\citep[LASCO/C2,][]{1995SoPh..162..357B} 
on board the Solar and Heliospheric Observatory\,\citep[SOHO,][]{1995SoPh..162....1D}, the Solar Ultraviolet Imager\,\citep[SUVI,][]{2018ApJ...852L...9S,2019SPIE11180E..7PV} 
on board the Geostationary Operational Environmental Satellite 17 (GOES-17) spacecraft and the Atmospheric Imaging Assembly\,\citep[AIA,][]{2012SoPh..275...17L} 
on board the Solar Dynamics Observatory\,\citep[SDO,][]{2012SoPh..275....3P}, \citet{2024A&A...683A.126L} recently studied the origin of a streamer blob.
They found that the streamer blob was formed by the gradual merging of three bright clumps initiated from the lower corona at about 1.8\,\rsun, 
which was driven by the expansion of the loop system at the base.
Their study indicates that the activity in the coronal base of a streamer might 
strongly affect the properties of the blobs formed above. The study of \citet{2024ApJ...973..130A} confirms that, within the simple magnetic environment of solar minimum, small-scale disturbances originate in the low corona($\sim$1.6\,\rsun) and evolve along nonradial paths into two distinct classes of outflows with different velocities.
On the other hand, there are two types of streamers grouped based on the activity at their bases: active region streamers (ARSs), 
and quiescent equatorial streamers (QESs).
An ARS lies above active regions, while a QES  has no underlying active regions, but prominences are usually present\,\citep[e.g.][]{1997SoPh..175..645R,1998SSRv...85..283R, 2003ApJ...585.1062U}.
\citet{1997SoPh..175..645R} found that the chemical abundance and The first ionization potential (FIP) effects in these two types of streamers can be significantly different.
The electron density in an ARS is higher than that in a QES\,\citep{2003ApJ...593.1146B}.
By contrast, the temperature of an ARS is normally lower than that of a QES\,\citep{2003MmSAI..74..717P}.
It remains an unanswered question whether the activity at the base of a streamer affects the properties of blobs produced therein.
To answer this question, we carried out a statistical analysis of the properties of blobs 
(including heights of first appearance, velocity, and acceleration) originating in these two different types of streamers (ARS and QES).
The rest of the paper is organized as follows: the data and method are described 
in Sect.~\ref{sec:observation}; the results are reported in Sect. ~\ref{sec:results}, 
and the discussion and conclusions are given in Sect.\ref{sec:conclusion}.

\begin{figure*}
\sidecaption
\includegraphics[width=12cm]{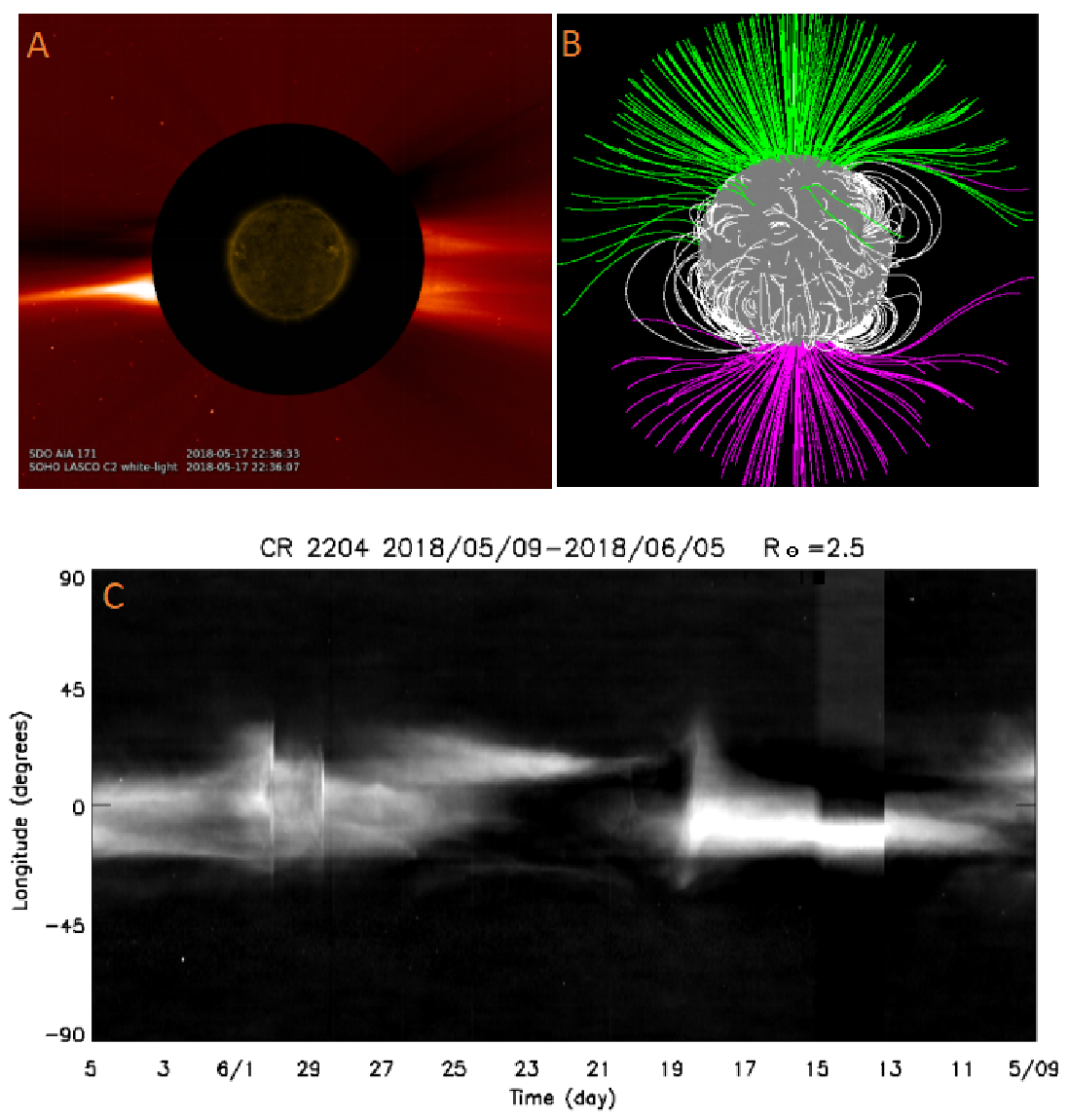}
\caption{Example of the streamer classification. Panel~A shows a composite image combining LASCO/C2 and AIA~171\,\AA\ images on 17 May 2018, from which the evolution in these observations allows us to determine whether there an active region exists at the streamer base.
The streamer shown at the east limb (left side of the image) is an example of an ARS located above the active region of NOAA~12711 (as seen in the AIA~171 \AA~image). 
Panel~B shows the PFSS model closest in time to the same day, which helps us to confirm that the streamer is a helmet streamer.
Panel~C is the Carrington map of solar rotation 2204, which is used to determine the time range of this helmet streamer (10-19  May 2018 for this case).}
\label{fig1}
\end{figure*}

\begin{figure*}[!ht]
\centering
\includegraphics[width=\textwidth]{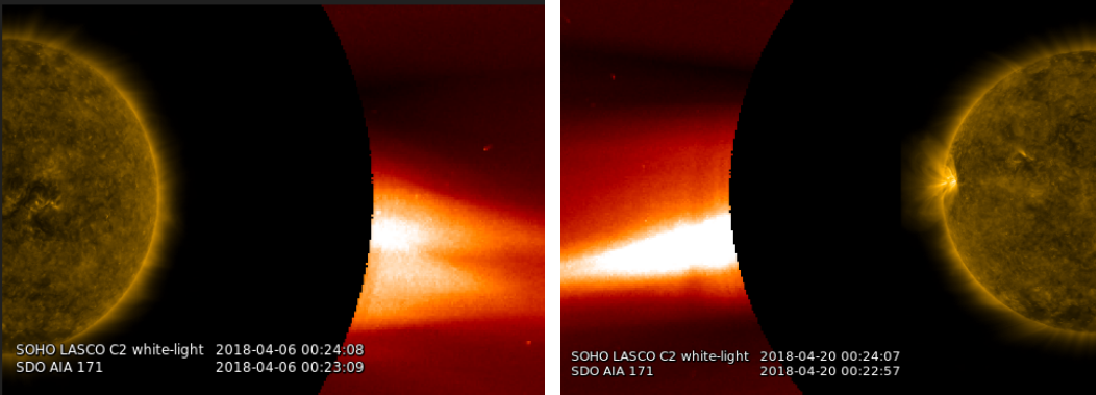}
\centering
\caption{LASCO/C2 and  AIA~171~\AA\ composite images of a QES (left panel) and an ARS (right panel).
In the left panel, the QES is visible at the west limb, where no active region is visible at its base throughout its lifetime (see the AIA~171~\AA~observations). 
In the right panel, the ARS is visible at the east limb, with an active region (NOAA~12706) at its base. All times denoted in the figures refer to universal time (UT).}
\label{fig2}
\end{figure*}

\begin{figure*}[!ht]
\centering
\includegraphics[width=\textwidth]{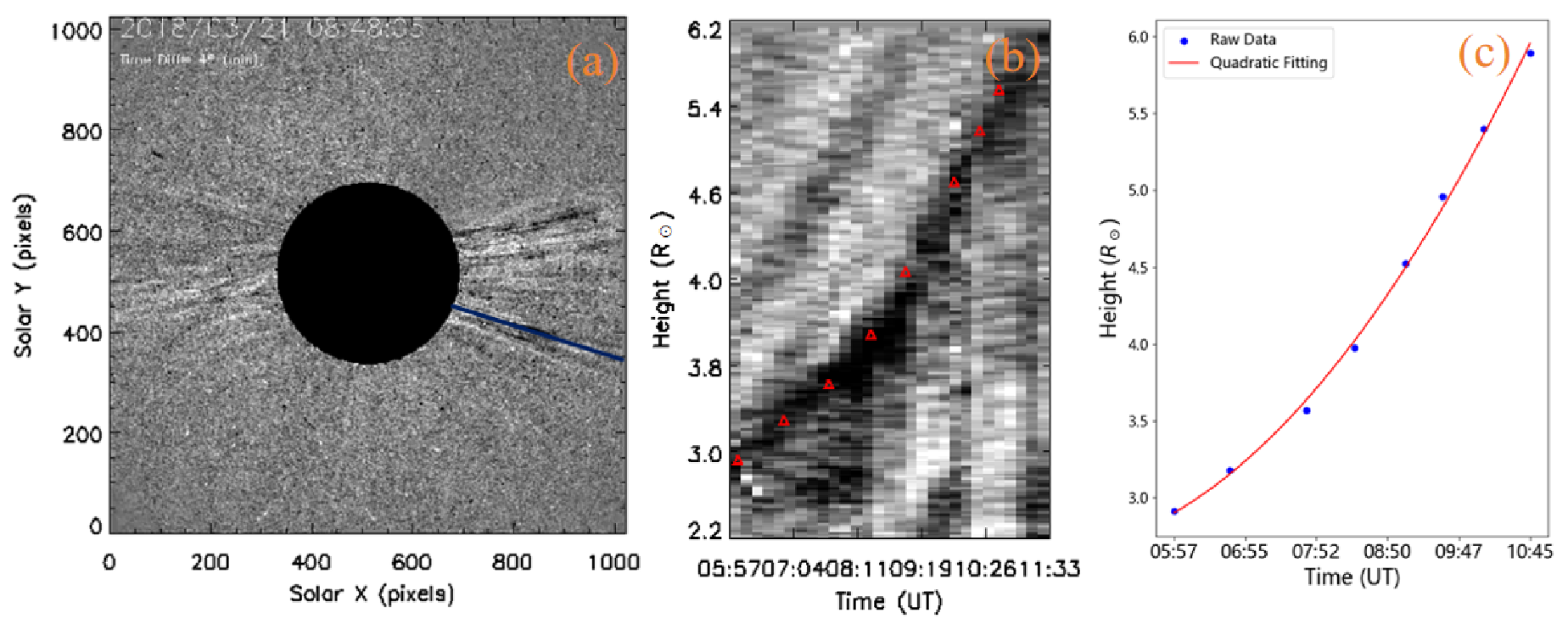}
\centering
\caption{Example showing how we obtained the parameters of a streamer blob.
Panel~(a) shows a running-difference image of LASCO/C2, where the dark blue line marks the propagating path of a streamer blob.
Panel~(b) is a time-distance diagram obtained along the dark blue line marked in panel (a) based on the running-difference images. 
The propagation of the blob is shown as the dark feature and the bright one above, marked by red triangles.
Based on the trajectory of the blob shown in panel~(b), a fitting curve was obtained using the quadratic fitting method, as shown in panel~(c).}
\label{fig3}
\end{figure*}

\begin{figure*}
\sidecaption
\includegraphics[width=12cm]{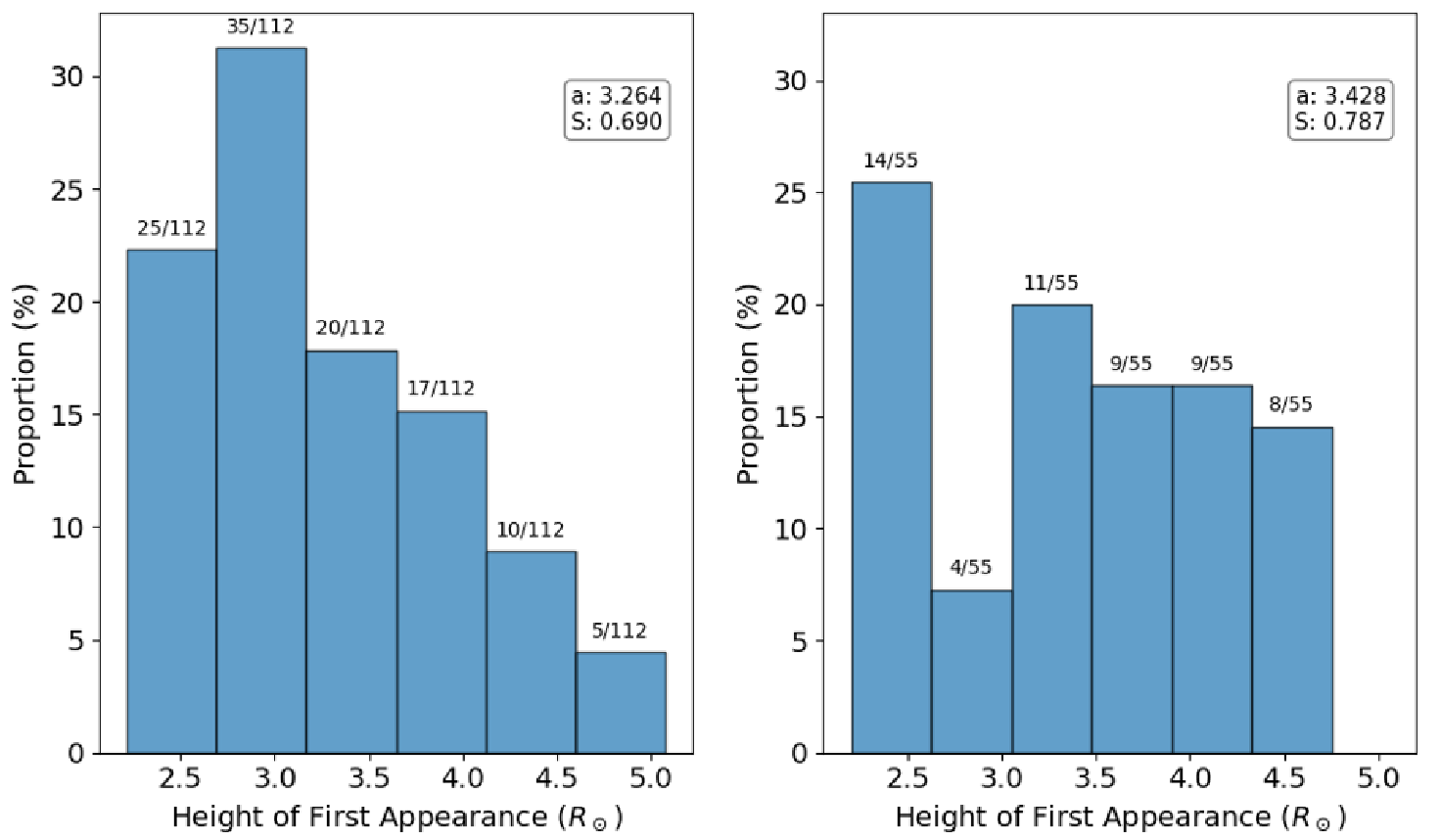} 
\caption{Statistical results of the heights of first appearance of ARS (left panel) and QES blobs (right panel).
The average and the standard deviation of the distributions are denoted a and s, respectively.}
\label{fig4}
\end{figure*}

\begin{figure*}
\sidecaption
\includegraphics[width=12cm]{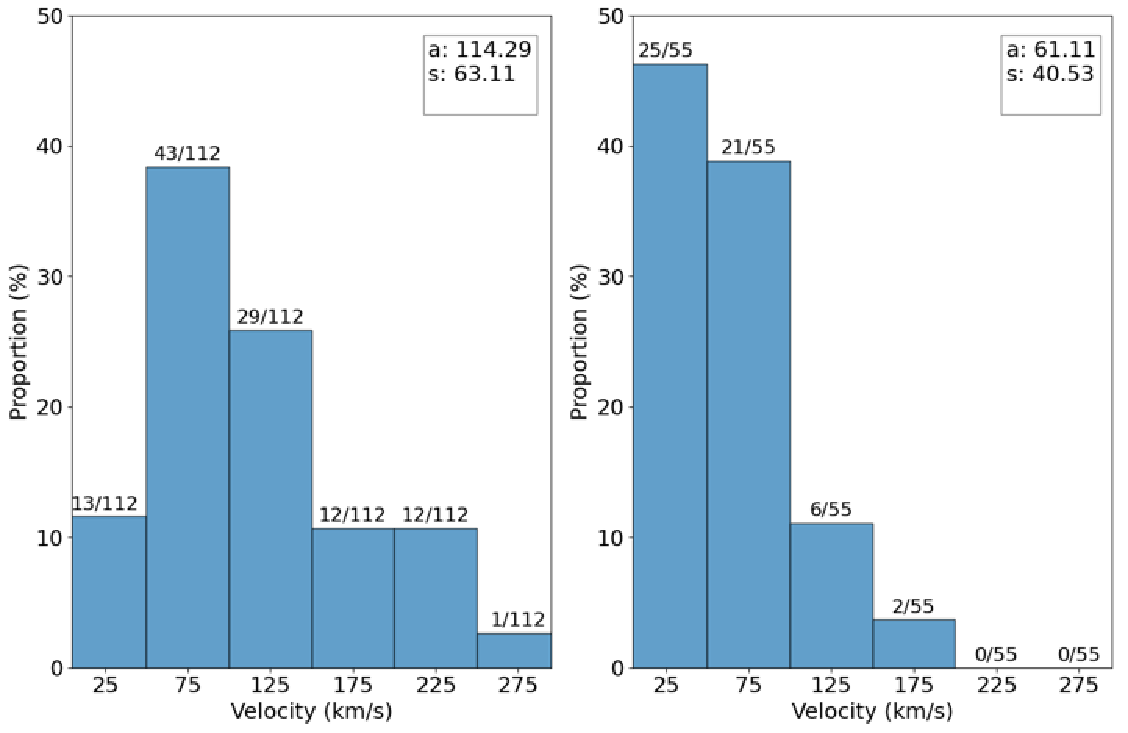} 
\caption{Same as Fig.\,\ref{fig4}, but for initial velocities of the blobs in two distinct streamer groups.}
\label{fig5}
\end{figure*}

\begin{figure*}
\sidecaption
\includegraphics[width=12cm]{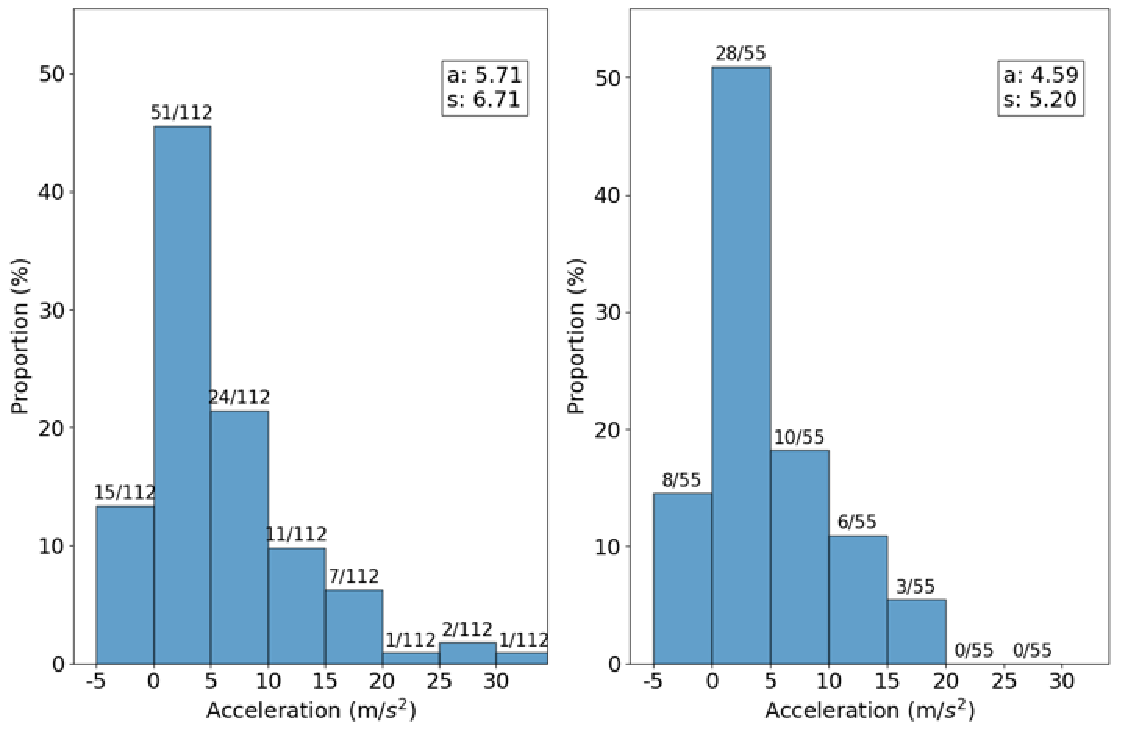} 
\caption{Same as Fig.\,\ref{fig4}, but for accelerations of the blobs in two distinct streamer groups.}
\label{fig6}
\end{figure*}

\begin{figure*}
\sidecaption
\includegraphics[width=12cm]{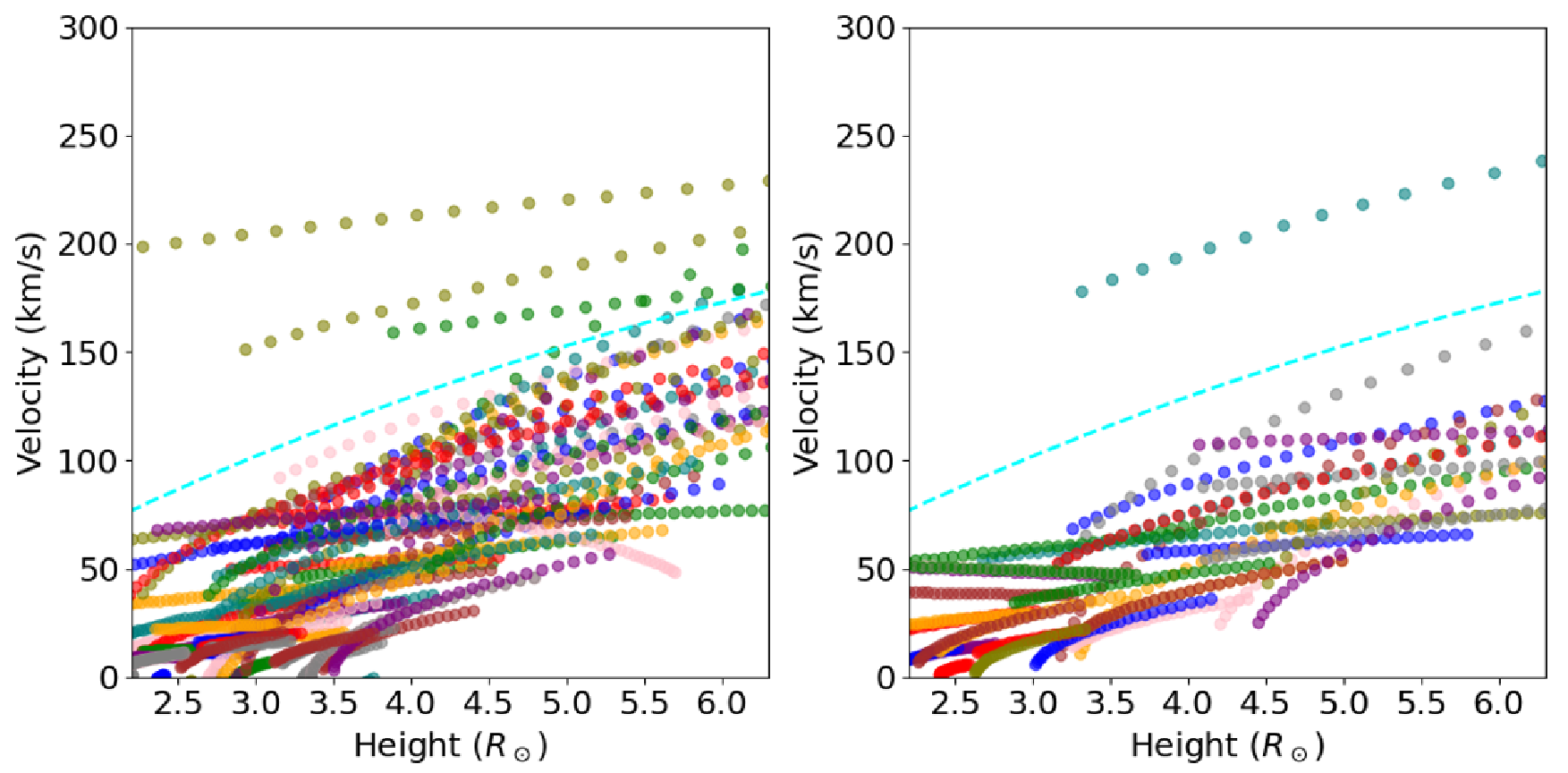}
\caption{Velocity as a function of height for 112 individual ARS blobs (left panel) and for 55 QES blobs (right panel).
Each individual blob is traced by a solid circle of the same color. 
The cyan lines represent the solar wind speeds given by the Paker model\,\citep{1958ApJ...128..664P}.}
\label{fig7}
\end{figure*}

\begin{figure*}
\sidecaption
\includegraphics[width=12cm]{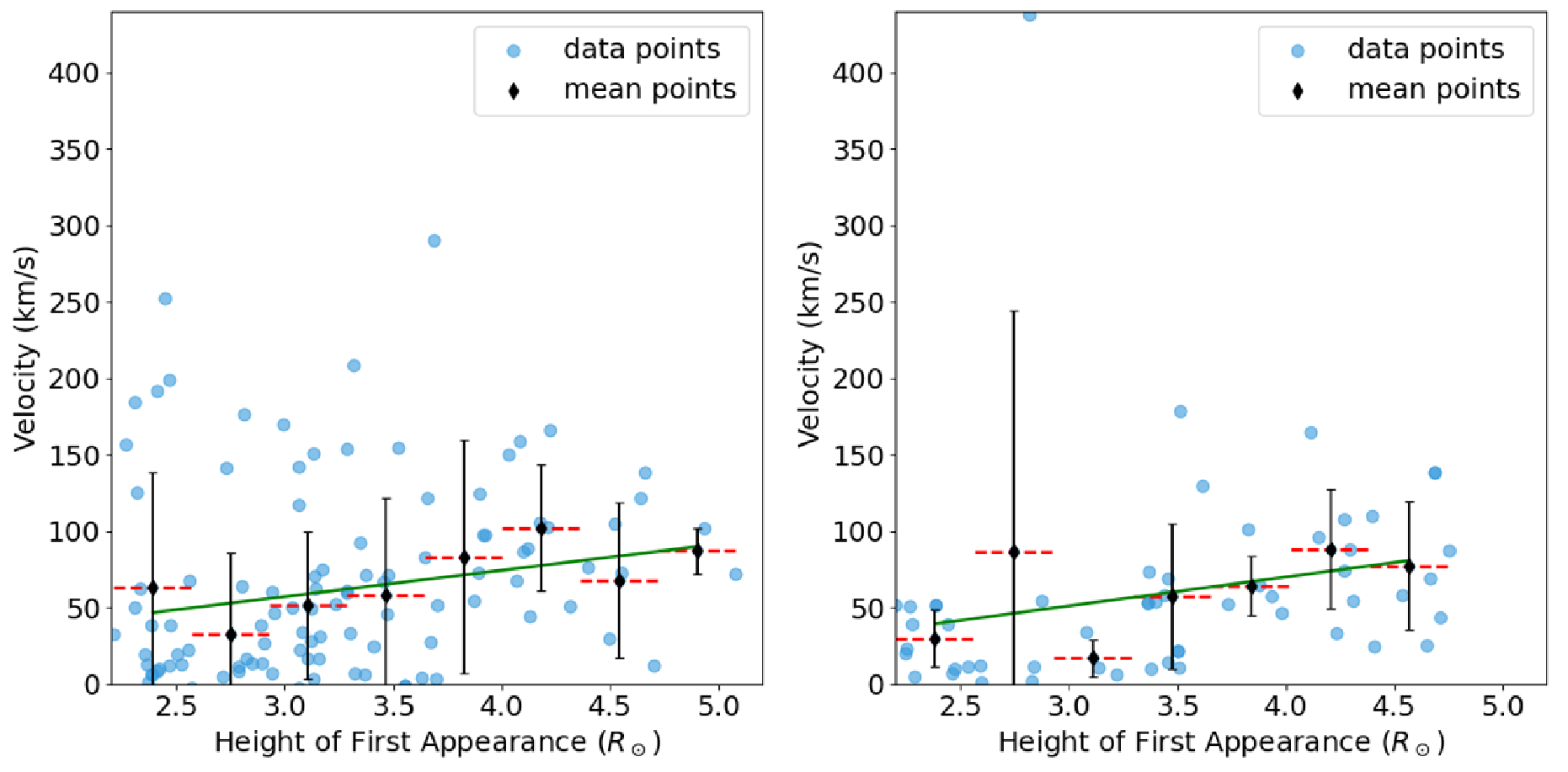}
\caption{Scatter diagrams for initial velocities vs. heights of first appearance of the ARS blobs (left panel) and of the QES blobs (right panel).
Each blue circle represents an individual blob.
A black diamond shows the average value obtained in a segment of 0.4\,\rsun\ (marked by the dashed red line), and the error bar in black denotes the 1$\sigma$ error.
The solid green line shows a linear fit to the segmented average values vs. height, which gives $y=17.2x+5.7$ for the left panel and $y=18.9x-5.5$ for the right panel.
 }
\label{fig8}
\end{figure*}

\begin{figure*}
\sidecaption
\includegraphics[width=12cm]{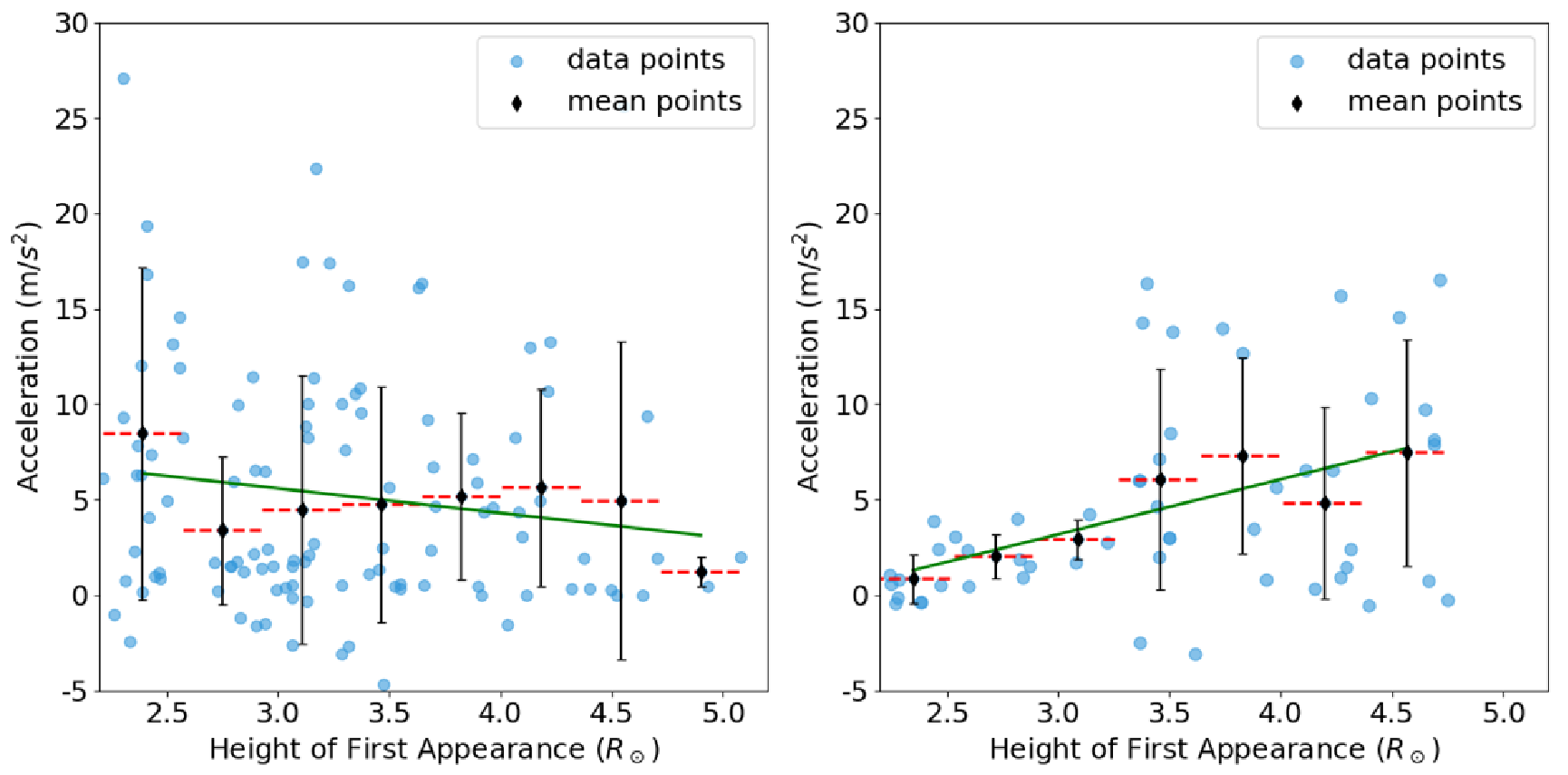}
\centering
\caption{Same as Fig.\,\ref{fig8}, but for the relation between accelerations and heights of first appearance.
The linear fits give $y=-1.3x+9.5$ and $y=2.9x-4.9$ for the left and right panels, respectively.}
\label{fig9}
\end{figure*}

\section{Observations and method}
\label{sec:observation}
The data analyzed here were taken by LASCO/C2, which is an externally occulted white-light coronal imager designed 
with a near 360\deg field of view (FOV) with the inner boundary at 2.2\,\rsun\ and the outer boundary at 6.5\,\rsun\,\citep{2025SoPh..300....4V}.
The images have a pixel size of 11.4\arcsec\ and a cadence of 45~s. 

\par
Data collected throughout 2018 were used for the study. They cover the solar minimum, which helps
us to avoid potential effects from large-scale eruptions, such as coronal mass ejections.
The data were calibrated by standard procedures provided by the instrument team, along with the solarsoft package, and
the time-independent emission from the F corona was removed by subtracting a monthly averaged background.

\par
To identify bright streamers and distinguish the conditions at their bases, we combined
AIA~171~\AA\ images, the coronal field extrapolations based on potential field source surface (PFSS), 
and Carrington-format LASCO/C2 maps.
In Fig.~\ref{fig1} we show an example of the classification principle.
A streamer was identified and followed in the LASCO/C2 images (Fig.\,\ref{fig1}A), and the PFSS model was used to confirm that 
it is a helmet streamer and not a pseudo-streamer (Fig.\,\ref{fig1}B).
The lifetime of the streamer was determined from the Carrington-format LASCO/C2 maps (Fig.\,\ref{fig1}C).
We then used images of the lower corona taken with the AIA~171\,\AA\ filter over the lifetime of the streamer to determine its category. 
When the base of a streamer had no active region during its full lifetime, it was considered a QES (see an example in the left panel of Fig.\,\ref{fig2}). 
When an active region was found in the base, it was considered an ARS  (see an example in the right panel of Fig.\,\ref{fig2}).

\par
In the LASCO/C2 observations taken throughout 2018, we followed 35 helmet streamers in total, including 17 ARSs and 18 QESs, and we searched for propagating blobs throughout their lifetimes.
Running-difference images based on LASCO/C2 observations were produced in order to highlight the faint moving blobs in the streamers, 
from which we were able to derive their properties.
In 17~ARSs, 112 blobs were found, and in 18~QESs, 55~blobs were found.
The occurrence rate of blobs in ARSs is about twice that in QESs, but the selection effect that 
some blobs are too faint to be found has to be taken into account, especially in QESs.
In Fig.\,\ref{fig3} we show an example of the analysis of a streamer blob.
Each streamer blob was identified from the time series of the LASCO/C2 running-difference images (Fig.\,\ref{fig3}(a)).
Along its propagation trajectory, the height--time (H--T) map was produced (Fig.\,\ref{fig3}(b)), 
from which its propagation can be tracked by the pair of dark and bright stripes (see the red triangles in Fig.\,\ref{fig3}(b)) and its heights as a function of time can be obtained (Fig.\,\ref{fig3}(c)).
We then fit its variation of heights versus time by a quadratic function\,\citep{1997ApJ...484..472S}, 
from which we obtained its height and velocity at the time that it was first observed in LASCO C2 (namely, heights of first appearance and initial velocity, respectively) 
and its acceleration in the FOV.
The same procedure was applied to each streamer blob, and the aforementioned properties were analyzed statistically.

\section{Results}
\label{sec:results}

\subsection{ Height of first appearance}
We defined the initial position of a tracked blob as the height of first appearance within the LASCO C2 field of view, from which the blob starts to show as a continuous track in the time-height map. 
It is important to note that due to the occulting disk ($< $2.2\,\rsun) and signal-to-noise ratio (S/N) limitations, the exact physical reconnection site (i.e., the pinch-off point \citep{2018ApJ...859..135W}) is often unresolved.
In all events, only nine of the ARS blobs and four of the QES blobs (7.8\% of the total number of samples) show signatures of pinch-off in the observations.
In nine ARS blobs, the pinch-off locations are in the range of 2.8$\sim$3.8\,\rsun, compared to 2.6$\sim$4.0\,\rsun\ in four QES blobs.
This indicates that the pinch-off process might take place at locations higher than 2.0\,\rsun\, but the probability is low.

\par
In Fig.\,\ref{fig4} we show the statistical results of the height of first appearance of the blobs in ARSs (left panel) and QESs (right panel).
All the studied streamer blobs first appear at heights ranging from 2.2\,\rsun\ to 5.1 \rsun.
With more samples, the statistical results here further expand on the previous findings of \citet{1997ApJ...484..472S}, who found that streamer blobs originate at about 3--4\,\rsun. On average, the heights of first appearance of blobs in ARSs are {3.3}$\pm${0.7}\,\rsun, and those in QESs are {3.4}$\pm${0.8}\,\rsun. 
Compared to ARS blobs, the distribution of QES blobs is flatter, and its peak is less significant.
This result shows that there is no significant difference in the heights of first appearance between ARS and QES blobs.
We would like to note that these numbers might be affected by a visibility effect due to the instrument sensitivity.

\subsection{Initial velocities and accelerations}
The distributions of initial velocities of blobs in the two types of streamers are shown in Fig.\,\ref{fig5}.
The distribution of ARS blobs peaks at about 75\,\kms, while that of QES blobs peaks at about 25\,\kms.
The average initial velocity of blobs in ARSs is approximately 114.29\,\kms, and it is higher than 61.11\,\kms\ in QESs.
The blob distribution in ARSs is wider than that in the QESs, with standard deviations of 63.11\,\kms\ in ARSs and 40.53\,\kms\ in QESs.
About 50\% of the blobs in ARSs compared with about 16\% of blobs in QESs have initial velocities higher than 100\,\kms, 
which is a typical coronal sound speed around these heights\,\citep{1988sscd.conf..130S}.
In both groups of blobs, just a few might exceed the local coronal Alfv\'en speed \citep[more than 200\,\kms see][]{1988sscd.conf..130S}).

\par
The distributions of the blob accelerations are shown in Fig.\,\ref{fig6}.
While the blobs propagate outward, most of them accelerate, as shown in the LASCO/C2 FOV.
Only 15 out of 112 ARS blobs and 8 out of 55 QES blobs decelerate.
Clearly, the distributions of ARS and QES blobs peak at about 2.5\,\mss.
On average, the accelerations of ARS blobs are 5.71$\pm$6.71\,\mss and those of QES blobs are 4.59$\pm$5.20\,\mss.
Although the average acceleration of ARS blobs is greater than that of QES blobs, the two distributions overlap widely.

\par
In Fig.\,\ref{fig7} we show the velocities as a function of height for each blob in ARSs (left panel) and in QESs (right panel) based on a second-order polynomial fit.
The blob velocities in both types of coronal streamers at any particular heights are generally 
lower than the velocity predicted by the Parker model of the slow solar wind (dashed cyan lines in Fig.~\ref{fig7}). 
However, the scatter plots of their velocities are tightly clustered around a parabolic trajectory. 
This indicates that these features are passive tracers carried by the slow solar wind and not actively driven eruptive structures. This is consistent with previous studies\,\citep{1997ApJ...484..472S}.

\subsection{Relation of the streamer blob properties}
In Fig.\,\ref{fig8} we show the scatter plots of initial velocities versus heights of first appearance of the ARS blobs (left panel) and the QES blobs (right panel).
Although the distributions of the data points seem to be random, an average variation based on bins of 0.4\,\rsun\ shows a clear trend: the initial velocity of a blob tends to be higher at higher locations.
This is valid for both ARS and QES blobs (see the black diamonds and the green fitting lines in Figure\,\ref{fig8}).
Again, we point out that this relation is an average effect, and initial blob velocities at any particular height can span a wide range.

\par
In Fig.\,\ref{fig9} we show the scatter plot of accelerations versus heights of first appearance of the ARS blobs (left panel) and the QES blobs (right panel).
Similarly, the ranges of the spread around any particular height are quite wide, especially in the case of the ARS blobs. 
For ARS blobs, the bin averages show a weak negative correlation between the accelerations and height of first appearance (see the left panel of Fig.\,\ref{fig9}).
In contrast, those for QES blobs show a clear positive correlation (see the right panel of Fig.\,\ref{fig9}). 

\section{Conclusions and discussion}
\label{sec:conclusion}
Based on observations of LASCO/C2, we carried out a statistical analysis of the height of first appearance, initial velocity, and 
acceleration of plasma blobs in two distinct groups of coronal streamers, those with an active region in the base (ARS blobs) and those with quiet bases (QES blobs), in order to answer the question of whether the activity at the base of a streamer affects the properties of the blobs that originated above.
The main results are listed below. \\
(1) The occurrence rate of blobs in ARSs appears to be twice that in QESs.\\
(2) The ARS blobs are more diverse than QES blobs.\\
(3) The heights of first appearance of ARS blobs are slightly lower than but very similar to those of QES blobs.
ARS blobs are {3.2}$\pm${0.7}\,\rsun\ and concentrated within the range of {2.2}-{4.0}\,\rsun, in contrast to QES blobs at  {3.4}$\pm${0.8}\,\rsun\,, whose distribution is relatively flatter and dispersed. \\
(4) The initial velocities of ARS blobs are 114.29$\pm$63.11\,\kms, which is much higher than the initial velocities of 61.11$\pm$40.53\,\kms\ for QES blobs.\\
(5) The initial velocities of about 50\% of the ARS blobs and of only about 16\% of the QES blobs are higher than the local sound speed, 
but very few blobs can exceed the local coronal Alfv\'en speed.\\
(6) The accelerations of ARS blobs are 5.71$\pm$6.71\,\mss, which is similar to 4.59$\pm$5.20\,\mss\ for QES blobs.\\
(7) For both groups, the variation in the velocities as a function of height of almost all the blobs can be well fit by second-order polynomial functions, 
but they are normally slower than the predictions of Parker's solar wind model.\\
(8) For both groups, the correlations between the initial velocities and heights of first appearance are relatively weak.\\
(9) The accelerations and heights of first appearance in ARS blobs are weakly negative-correlated, but a clear positive correlation is present in QES blobs.

\par
Previous studies have shown that various activities in the lower corona might be the direct or indirect origin of the solar wind in 
planetary space\,\citep{2021NatAs...5.1029S,2023NatAs...7..133C,2023SoPh..298...78W}.
The origin of streamer blobs can also be driven by activities from below\,\citep{2021ApJ...920L...6L,2024A&A...683A.126L}.
While \citet{2024ApJ...973..130A} focused on a statistical study of transients within a single streamer, we performed a statistics analysis of different streamer environments to show that the mechanisms for producing these transients might vary depending on the environment characteristics.
The results we obtained confirm that activities at the base of a streamer can significantly affect the properties of blobs that originate above it.
The origin of a streamer blob from interchange magnetic reconnection\,\citep{1998ApJ...498L.165W}, a pinching-off process\,\citep{2018ApJ...859....6H, 2024A&A...683A.126L}, or a tearing-mode instability at loop tops\,\citep{1998ApJ...495..491K,1999JGR...104..521E} is thought to require a driver of loop expansions from below.
Therefore, ARS blobs are expected to start from a lower height than QES blobs because loops in a single active region are normally smaller and thus lower than those crossing a quiet region and connecting two different active regions\,\citep{2014LRSP...11....4R}.
This might explain why the mean height of first appearance of ARS blobs (3.2\,\rsun) is lower than that of QES blobs (3.4\,\rsun).
However, a visibility effect might also play a role.
As deduced by \citet{2018ApJ...859..135W}, the visibility of these newly formed flux ropes depends on the amount of material they sweep up. 
Because ARS streamers are subject to enhanced heating and are inherently denser and brighter, the emerging blobs reach the observational contrast threshold at lower altitudes. 
In contrast, in the more tenuous QES environment, blobs must typically propagate to greater heights to accumulate sufficient material to become detectable against the background. 
Thus, the lower first-appearance height in ARS reflects their stronger underlying dynamics and not a lower physical pinch-off site.

\par
The initial velocity of a streamer blob is likely related to its generation processes.
The higher initial velocities in ARS blobs suggest a more violent start than for QES blobs.
This is reasonable, since the active region beneath an ARS is much more dynamic than the quiet region beneath a QES. 
Weak positive correlations between initial velocities and heights of first appearance are present in these blobs, 
and this is inconsistent with the variations in Alfv\'en speeds in this region, which are expected to decrease with increasing height \,\citep{2005A&A...435.1123W}.
This suggests that the origin of these blobs is diverse, and magnetic reconnection might not be the only mechanism.
The acceleration of a streamer blob is very similar to that of the ambient solar wind, indicating that it is picked up by the ambient solar wind soon after its birth, and thus becomes a source of micro-streams in the solar wind.

\par
In these two groups of streamer blobs, the variation in accelerations as a function of height appears to be different from one blob to the next.
In ARS blobs, the negative correlation between the accelerations and heights of first appearance might suggest a strong effect from the magnetic field strength, which decreases with height.
The opposite correlation shown in QES blobs suggests that the ambient solar wind might play a major role in the acceleration.
At higher altitudes, the difference between the solar wind speed and the blobs is larger, potentially providing greater acceleration.
This might be a plausible explanation for this result, but the reality might be much more complicated.

\par
To summarize, we compared the properties of propagating blobs in two different groups of streamers 
(ARS and QES) that have different activities at their bases.
Although in general, the dynamics of a streamer blob can be different from one to another, 
we provided statistical evidence that the activities at the base of a streamer can affect the dynamics of a blob originating above.
Therefore, we infer that the dynamics in the lower corona might be crucial in generating small-scale structures in the solar wind.

\begin{acknowledgements}
{We are grateful to the anonymous referee for their critical comments that improved the manuscript.
This research is supported by the National Key R\&D Program of China No. 2021YFA0718600, 
National Natural Science Foundation of China (42230203, 42174201,42474223) and China's Space Origins Exploration Program.
M.M. acknowledges DFG grant WI 3211/8-2, project number 452856778.
Z.H. and M.M. acknowledge the support by the International Space Science Institute (ISSI) in Bern, through ISSI International Team project \#24-605 (Small-scale magnetic flux ropes under the microscope with Parker Solar Probe and Solar Orbiter).
M.M. thanks the International Space Science Institute, Bern, Switzerland, for the Visiting Scientist grant.
We thank Dr. Binbin Tang for the valuable discussion on the theory of interchange magnetic reconnection.
SOHO is a project of international cooperation between ESA and NASA.
The SOHO/LASCO data used here are produced by a consortium of the Naval Research Laboratory (USA), 
Max-Planck Institute for Sonnensystemforschung (Germany), Laboratoire d'Astrophysique de Marseille (France), and the University of Birmingham (UK).
The AIA and HMI data are used by courtesy of NASA/SDO, the AIA and HMI teams, JSOC, and the Helioviewer team.}
\end{acknowledgements}


\bibliographystyle{aa}
\bibliography{streamer_paper}

\begin{thebibliography}{52}
\expandafter\ifx\csname natexlab\endcsname\relax\def\natexlab#1{#1}\fi

\bibitem[{{Alzate} {et~al.}(2024){Alzate}, {Di Matteo}, {Morgan}, {Viall}, \&
  {Vourlidas}}]{2024ApJ...973..130A}
{Alzate}, N., {Di Matteo}, S., {Morgan}, H., {Viall}, N., \& {Vourlidas}, A.
  2024, \apj, 973, 130

\bibitem[{{Ambro{\v{z}}} {et~al.}(2009){Ambro{\v{z}}}, {Druckm{\"u}ller},
  {Galal}, \& {Hamid}}]{2009SoPh..258..243A}
{Ambro{\v{z}}}, P., {Druckm{\"u}ller}, M., {Galal}, A.~A., \& {Hamid}, R.~H.
  2009, \solphys, 258, 243

\bibitem[{{Antonucci} {et~al.}(2023){Antonucci}, {Downs}, {Capuano}, {Spadaro},
  {Susino}, {Telloni}, {Andretta}, {Da Deppo}, {De Leo}, {Fineschi},
  {Frassetto}, {Landini}, {Naletto}, {Nicolini}, {Pancrazzi}, {Romoli},
  {Stangalini}, {Teriaca}, \& {Uslenghi}}]{2023PhPl...30b2905A}
{Antonucci}, E., {Downs}, C., {Capuano}, G.~E., {et~al.} 2023, Physics of
  Plasmas, 30, 022905

\bibitem[{{Bemporad} {et~al.}(2003){Bemporad}, {Poletto}, {Suess}, {Ko},
  {Parenti}, {Riley}, {Romoli}, \& {Zurbuchen}}]{2003ApJ...593.1146B}
{Bemporad}, A., {Poletto}, G., {Suess}, S.~T., {et~al.} 2003, \apj, 593, 1146

\bibitem[{{Boe} {et~al.}(2020){Boe}, {Habbal}, \&
  {Druckm{\"u}ller}}]{2020ApJ...895..123B}
{Boe}, B., {Habbal}, S., \& {Druckm{\"u}ller}, M. 2020, \apj, 895, 123

\bibitem[{{Brueckner} {et~al.}(1995){Brueckner}, {Howard}, {Koomen},
  {Korendyke}, {Michels}, {Moses}, {Socker}, {Dere}, {Lamy}, {Llebaria},
  {Bout}, {Schwenn}, {Simnett}, {Bedford}, \& {Eyles}}]{1995SoPh..162..357B}
{Brueckner}, G.~E., {Howard}, R.~A., {Koomen}, M.~J., {et~al.} 1995, \solphys,
  162, 357

\bibitem[{{C{\'e}cere} {et~al.}(2025){C{\'e}cere}, {Wyper}, {Krause}, {Sahade},
  \& {Rice}}]{2025A&A...697A.155C}
{C{\'e}cere}, M., {Wyper}, P.~F., {Krause}, G., {Sahade}, A., \& {Rice},
  O.~E.~K. 2025, \aap, 697, A155

\bibitem[{{Chen} {et~al.}(2009){Chen}, {Li}, {Song}, {Shi}, {Feng}, \&
  {Xia}}]{2009ApJ...691.1936C}
{Chen}, Y., {Li}, X., {Song}, H.~Q., {et~al.} 2009, \apj, 691, 1936

\bibitem[{{Chitta} {et~al.}(2023){Chitta}, {Seaton}, {Downs}, {DeForest}, \&
  {Higginson}}]{2023NatAs...7..133C}
{Chitta}, L.~P., {Seaton}, D.~B., {Downs}, C., {DeForest}, C.~E., \&
  {Higginson}, A.~K. 2023, Nature Astronomy, 7, 133

\bibitem[{{Domingo} {et~al.}(1995){Domingo}, {Fleck}, \&
  {Poland}}]{1995SoPh..162....1D}
{Domingo}, V., {Fleck}, B., \& {Poland}, A.~I. 1995, \solphys, 162, 1

\bibitem[{{Einaudi} {et~al.}(1999){Einaudi}, {Boncinelli}, {Dahlburg}, \&
  {Karpen}}]{1999JGR...104..521E}
{Einaudi}, G., {Boncinelli}, P., {Dahlburg}, R.~B., \& {Karpen}, J.~T. 1999,
  \jgr, 104, 521

\bibitem[{{Geiss} {et~al.}(1995){Geiss}, {Gloeckler}, \& {von
  Steiger}}]{1995SSRv...72...49G}
{Geiss}, J., {Gloeckler}, G., \& {von Steiger}, R. 1995, \ssr, 72, 49

\bibitem[{{Gosling} {et~al.}(1981){Gosling}, {Borrini}, {Asbridge}, {Bame},
  {Feldman}, \& {Hansen}}]{1981JGR....86.5438G}
{Gosling}, J.~T., {Borrini}, G., {Asbridge}, J.~R., {et~al.} 1981, \jgr, 86,
  5438

\bibitem[{{Higginson} \& {Lynch}(2018)}]{2018ApJ...859....6H}
{Higginson}, A.~K. \& {Lynch}, B.~J. 2018, \apj, 859, 6

\bibitem[{{Karpen} {et~al.}(1998){Karpen}, {Antiochos}, {Richard DeVore}, \&
  {Golub}}]{1998ApJ...495..491K}
{Karpen}, J.~T., {Antiochos}, S.~K., {Richard DeVore}, C., \& {Golub}, L. 1998,
  \apj, 495, 491

\bibitem[{{Koutchmy} \& {Livshits}(1992)}]{1992SSRv...61..393K}
{Koutchmy}, S. \& {Livshits}, M. 1992, \ssr, 61, 393

\bibitem[{{Lee} {et~al.}(2021){Lee}, {Cho}, {An}, {Lee}, {Seough}, {Kim}, \&
  {Kumar}}]{2021ApJ...920L...6L}
{Lee}, J.-O., {Cho}, K.-S., {An}, J., {et~al.} 2021, \apjl, 920, L6

\bibitem[{{Lemen} {et~al.}(2012){Lemen}, {Title}, {Akin}, {Boerner}, {Chou},
  {Drake}, {Duncan}, {Edwards}, {Friedlaender}, {Heyman}, {Hurlburt}, {Katz},
  {Kushner}, {Levay}, {Lindgren}, {Mathur}, {McFeaters}, {Mitchell}, {Rehse},
  {Schrijver}, {Springer}, {Stern}, {Tarbell}, {Wuelser}, {Wolfson}, {Yanari},
  {Bookbinder}, {Cheimets}, {Caldwell}, {Deluca}, {Gates}, {Golub}, {Park},
  {Podgorski}, {Bush}, {Scherrer}, {Gummin}, {Smith}, {Auker}, {Jerram},
  {Pool}, {Soufli}, {Windt}, {Beardsley}, {Clapp}, {Lang}, \&
  {Waltham}}]{2012SoPh..275...17L}
{Lemen}, J.~R., {Title}, A.~M., {Akin}, D.~J., {et~al.} 2012, \solphys, 275, 17

\bibitem[{{Li} {et~al.}(2024){Li}, {Huang}, {Deng}, {Fu}, {Xia}, {Song},
  {Xiong}, {Wei}, {Qi}, \& {Zhang}}]{2024A&A...683A.126L}
{Li}, H., {Huang}, Z., {Deng}, K., {et~al.} 2024, \aap, 683, A126

\bibitem[{{Liang} {et~al.}(2023){Liang}, {Qu}, {Hao}, {Xu}, \&
  {Zhong}}]{2023MNRAS.518.1776L}
{Liang}, Y., {Qu}, Z., {Hao}, L., {Xu}, Z., \& {Zhong}, Y. 2023, \mnras, 518,
  1776

\bibitem[{{Lynch}(2020)}]{2020ApJ...905..139L}
{Lynch}, B.~J. 2020, \apj, 905, 139

\bibitem[{{Lyu} {et~al.}(2023){Lyu}, {Wang}, {Li}, \&
  {Zhang}}]{2023A&A...672A.100L}
{Lyu}, S., {Wang}, Y., {Li}, X., \& {Zhang}, Q. 2023, \aap, 672, A100

\bibitem[{{Lyu} {et~al.}(2025){Lyu}, {Wang}, \& {Owen}}]{2025ApJ...988..152L}
{Lyu}, S., {Wang}, Y., \& {Owen}, C.~J. 2025, \apj, 988, 152

\bibitem[{{Parenti} {et~al.}(2003){Parenti}, {Landi}, \&
  {Bromage}}]{2003MmSAI..74..717P}
{Parenti}, S., {Landi}, E., \& {Bromage}, B.~J.~I. 2003, \memsai, 74, 717

\bibitem[{{Parker}(1958)}]{1958ApJ...128..664P}
{Parker}, E.~N. 1958, \apj, 128, 664

\bibitem[{{Pasachoff} {et~al.}(2011){Pasachoff}, {Ru{\v{s}}in},
  {Druckm{\"u}llerov{\'a}}, {Saniga}, {Lu}, {Malamut}, {Seaton}, {Golub},
  {Engell}, {Hill}, \& {Lucas}}]{2011ApJ...734..114P}
{Pasachoff}, J.~M., {Ru{\v{s}}in}, V., {Druckm{\"u}llerov{\'a}}, H., {et~al.}
  2011, \apj, 734, 114

\bibitem[{{Pasachoff} {et~al.}(2015){Pasachoff}, {Ru{\v{s}}in}, {Saniga},
  {Babcock}, {Lu}, {Davis}, {Dantowitz}, {Gaintatzis}, {Seiradakis},
  {Voulgaris}, {Seaton}, \& {Shiota}}]{2015ApJ...800...90P}
{Pasachoff}, J.~M., {Ru{\v{s}}in}, V., {Saniga}, M., {et~al.} 2015, \apj, 800,
  90

\bibitem[{{Pesnell} {et~al.}(2012){Pesnell}, {Thompson}, \&
  {Chamberlin}}]{2012SoPh..275....3P}
{Pesnell}, W.~D., {Thompson}, B.~J., \& {Chamberlin}, P.~C. 2012, \solphys,
  275, 3

\bibitem[{{Raymond} {et~al.}(1997){Raymond}, {Kohl}, {Noci}, {Antonucci},
  {Tondello}, {Huber}, {Gardner}, {Nicolosi}, {Fineschi}, {Romoli}, {Spadaro},
  {Siegmund}, {Benna}, {Ciaravella}, {Cranmer}, {Giordano}, {Karovska},
  {Martin}, {Michels}, {Modigliani}, {Naletto}, {Panasyuk}, {Pernechele},
  {Poletto}, {Smith}, {Suleiman}, \& {Strachan}}]{1997SoPh..175..645R}
{Raymond}, J.~C., {Kohl}, J.~L., {Noci}, G., {et~al.} 1997, \solphys, 175, 645

\bibitem[{{Raymond} {et~al.}(1998){Raymond}, {Suleiman}, {Kohl}, \&
  {Noci}}]{1998SSRv...85..283R}
{Raymond}, J.~C., {Suleiman}, R., {Kohl}, J.~L., \& {Noci}, G. 1998, \ssr, 85,
  283

\bibitem[{{Reale}(2014)}]{2014LRSP...11....4R}
{Reale}, F. 2014, Living Reviews in Solar Physics, 11, 4

\bibitem[{{R{\'e}ville} {et~al.}(2022){R{\'e}ville}, {Fargette}, {Rouillard},
  {Lavraud}, {Velli}, {Strugarek}, {Parenti}, {Brun}, {Shi}, {Kouloumvakos},
  {Poirier}, {Pinto}, {Louarn}, {Fedorov}, {Owen}, {G{\'e}not}, {Horbury},
  {Laker}, {O'Brien}, {Angelini}, {Fauchon-Jones}, \&
  {Kasper}}]{2022A&A...659A.110R}
{R{\'e}ville}, V., {Fargette}, N., {Rouillard}, A.~P., {et~al.} 2022, \aap,
  659, A110

\bibitem[{{Saito} \& {Tandberg-Hanssen}(1973)}]{1973SoPh...31..105S}
{Saito}, K. \& {Tandberg-Hanssen}, E. 1973, \solphys, 31, 105

\bibitem[{{Seaton} \& {Darnel}(2018)}]{2018ApJ...852L...9S}
{Seaton}, D.~B. \& {Darnel}, J.~M. 2018, \apjl, 852, L9

\bibitem[{{Seaton} {et~al.}(2021){Seaton}, {Hughes}, {Tadikonda}, {Caspi},
  {DeForest}, {Krimchansky}, {Hurlburt}, {Seguin}, \&
  {Slater}}]{2021NatAs...5.1029S}
{Seaton}, D.~B., {Hughes}, J.~M., {Tadikonda}, S.~K., {et~al.} 2021, Nature
  Astronomy, 5, 1029

\bibitem[{{Sheeley} {et~al.}(2009){Sheeley}, {Lee}, {Casto}, {Wang}, \&
  {Rich}}]{2009ApJ...694.1471S}
{Sheeley}, N.~R., J., {Lee}, D.~D.~H., {Casto}, K.~P., {Wang}, Y.~M., \&
  {Rich}, N.~B. 2009, \apj, 694, 1471

\bibitem[{{Sheeley} \& {Rouillard}(2010)}]{2010ApJ...715..300S}
{Sheeley}, N.~R., J. \& {Rouillard}, A.~P. 2010, \apj, 715, 300

\bibitem[{{Sheeley} {et~al.}(1997){Sheeley}, {Wang}, {Hawley}, {Brueckner},
  {Dere}, {Howard}, {Koomen}, {Korendyke}, {Michels}, {Paswaters}, {Socker},
  {St. Cyr}, {Wang}, {Lamy}, {Llebaria}, {Schwenn}, {Simnett}, {Plunkett}, \&
  {Biesecker}}]{1997ApJ...484..472S}
{Sheeley}, N.~R., {Wang}, Y.~M., {Hawley}, S.~H., {et~al.} 1997, \apj, 484, 472

\bibitem[{{Song} {et~al.}(2009){Song}, {Chen}, {Liu}, {Feng}, \&
  {Xia}}]{2009SoPh..258..129S}
{Song}, H.~Q., {Chen}, Y., {Liu}, K., {Feng}, S.~W., \& {Xia}, L.~D. 2009,
  \solphys, 258, 129

\bibitem[{{Song} {et~al.}(2012){Song}, {Kong}, {Chen}, {Li}, {Li}, {Feng}, \&
  {Xia}}]{2012SoPh..276..261S}
{Song}, H.~Q., {Kong}, X.~L., {Chen}, Y., {et~al.} 2012, \solphys, 276, 261

\bibitem[{{Spadaro} {et~al.}(2007){Spadaro}, {Susino}, {Ventura}, {Vourlidas},
  \& {Landi}}]{2007A&A...475..707S}
{Spadaro}, D., {Susino}, R., {Ventura}, R., {Vourlidas}, A., \& {Landi}, E.
  2007, \aap, 475, 707

\bibitem[{{Suess}(1988)}]{1988sscd.conf..130S}
{Suess}, S.~T. 1988, in Solar and Stellar Coronal Structure and Dynamics, ed.
  R.~C. {Altrock}, 130--139

\bibitem[{{Uzzo} {et~al.}(2003){Uzzo}, {Ko}, {Raymond}, {Wurz}, \&
  {Ipavich}}]{2003ApJ...585.1062U}
{Uzzo}, M., {Ko}, Y.~K., {Raymond}, J.~C., {Wurz}, P., \& {Ipavich}, F.~M.
  2003, \apj, 585, 1062

\bibitem[{{V{\'a}squez} {et~al.}(2025){V{\'a}squez}, {Nuevo}, {Burtovoi},
  {Lamy}, {Romoli}, {Gilardy}, {Frazin}, {Sachdeva}, {Manchester}, {Abbo}, {De
  Leo}, {Frassati}, {Jerse}, {Landini}, {Russano}, {Sasso}, {Susino}, \&
  {Uslenghi}}]{2025SoPh..300....4V}
{V{\'a}squez}, A.~M., {Nuevo}, F.~A., {Burtovoi}, A., {et~al.} 2025, \solphys,
  300, 4

\bibitem[{{Vasudevan} {et~al.}(2019){Vasudevan}, {Shing}, {Mathur}, {Edwards},
  {Shaw}, {Seaton}, \& {Darnel}}]{2019SPIE11180E..7PV}
{Vasudevan}, G., {Shing}, L., {Mathur}, D., {et~al.} 2019, in Society of
  Photo-Optical Instrumentation Engineers (SPIE) Conference Series, Vol. 11180,
  International Conference on Space Optics; ICSO 2018, ed. Z.~{Sodnik},
  N.~{Karafolas}, \& B.~{Cugny}, 111807P

\bibitem[{{Ventura} {et~al.}(2005){Ventura}, {Spadaro}, {Cimino}, \&
  {Romoli}}]{2005A&A...430..701V}
{Ventura}, R., {Spadaro}, D., {Cimino}, G., \& {Romoli}, M. 2005, \aap, 430,
  701

\bibitem[{{Wang} \& {Hess}(2018)}]{2018ApJ...859..135W}
{Wang}, Y.~M. \& {Hess}, P. 2018, \apj, 859, 135

\bibitem[{{Wang} \& {Sheeley}(1992)}]{1992ApJ...392..310W}
{Wang}, Y.~M. \& {Sheeley}, N.~R., J. 1992, \apj, 392, 310

\bibitem[{{Wang} {et~al.}(2000){Wang}, {Sheeley}, \&
  {Rich}}]{2000GeoRL..27..149W}
{Wang}, Y.~M., {Sheeley}, N.~R., J., \& {Rich}, N.~B. 2000, \grl, 27, 149

\bibitem[{{Wang} {et~al.}(1998){Wang}, {Sheeley}, {Walters}, {Brueckner},
  {Howard}, {Michels}, {Lamy}, {Schwenn}, \& {Simnett}}]{1998ApJ...498L.165W}
{Wang}, Y.~M., {Sheeley}, N.~R., J., {Walters}, J.~H., {et~al.} 1998, \apjl,
  498, L165

\bibitem[{{Warmuth} \& {Mann}(2005)}]{2005A&A...435.1123W}
{Warmuth}, A. \& {Mann}, G. 2005, \aap, 435, 1123

\bibitem[{{West} {et~al.}(2023){West}, {Seaton}, {Wexler}, {Raymond}, {Del
  Zanna}, {Rivera}, {Kobelski}, {Chen}, {DeForest}, {Golub}, {Caspi}, {Gilly},
  {Kooi}, {Meyer}, {Alterman}, {Alzate}, {Andretta}, {Auch{\`e}re}, {Banerjee},
  {Berghmans}, {Chamberlin}, {Chitta}, {Downs}, {Giordano}, {Harra},
  {Higginson}, {Howard}, {Kumar}, {Mason}, {Mason}, {Morton}, {Nykyri},
  {Patel}, {Rachmeler}, {Reardon}, {Reeves}, {Savage}, {Thompson}, {Van
  Kooten}, {Viall}, {Vourlidas}, \& {Zhukov}}]{2023SoPh..298...78W}
{West}, M.~J., {Seaton}, D.~B., {Wexler}, D.~B., {et~al.} 2023, \solphys, 298,
  78

\end{thebibliography}

\end{document}